\documentclass[pra, aps, twocolumn, groupedaddress, showpacs, superscriptaddress]{revtex4}
\usepackage{slashed}
\usepackage{graphicx}
\usepackage{subfigure}
\usepackage[usenames, dvipsnames]{color}
\usepackage{graphics,amsfonts,amsmath, amssymb}
\usepackage{hyperref}
\usepackage{bm}
\usepackage{amsfonts}
\usepackage{lipsum}

\hypersetup{backref,
colorlinks=true,
urlcolor=blue,
linkcolor=blue,
linktoc=page,
citecolor=blue}

\everymath{\displaystyle}

\begin{document}
\title{Phonon-mediated quantum discord in dark solitons}
\author{M. I. Shaukat}
\affiliation{ Instituto Superior T\'ecnico, University of Lisbon and Instituto de Telecomunica\c{c}\~{o}es, Torre Norte, Av. Rovisco Pais 1,
Lisbon, Portugal}
\affiliation{University of Engineering and Technology, Lahore (RCET Campus), Pakistan}
\email{muzzamalshaukat@gmail.com}
\author{A. Slaoui}
\affiliation{LPHE-Modeling and Simulation, Faculty of Sciences, University Mohammed V, Rabat, Morocco.}
\email{abdallah.slaoui@um5s.net.ma}
\author{H. Ter\c{c}as}
\affiliation{ Instituto Superior T\'ecnico and Instituto de Plasmas e Fus\~ao Nuclear,
Lisbon, Portugal}
\email{hugo.tercas@tecnico.ulisboa.pt}
\author{M. Daoud}
\affiliation{Department of Physics, Faculty of Sciences, University Ibn Tofail, K\'{e}nitra, Morocco}
\affiliation{Abdus Salam International Centre for Theoretical Physics, Miramare, Trieste, Italy.}


\begin{abstract}
    We investigate the quantum correlation dynamics in a dark-soliton qubits with special attention to quantum discord.     
    Recently, dark-soliton qubit exhibiting appreciably long lifetime are proved to be an excellent candidate for information processing.
     Depending on the precise distance between the dark-soliton qubits, the decay rate of Dicke symmetric and antisymmetric state is suppressed or enhanced.
     With the Renyi-2 entropy, we derive a simple analytical expression for the quantum discord, and explore the generation and decay of correlation for different initial states.
      We believe the present work could pave the stage for a new generation of quantum discord based purely on matter-wave phononics.

\end{abstract}

\maketitle

\section{Introduction}
Entanglement is a fundamental resource in quantum information (QI) and computation \cite{Nielsen, Ficek} that distinguishes the classical world from quantum one. One of the major challenges for the physical realization of QI protocols is the decoherence, resulting loss of information from the system to the environment. Due to the collective properties, the spontaneous emission of multi-atomic system can be altered in comparison to single atom case. Two different spontaneous decay rates (superradiant and subradiant) due to inter-atomic dipole interaction has been recognized by Dicke \cite{Dicke1954}, whereas Tavis et al. \cite{Tavis1968} studied the single-mode field interaction inside a cavity, as a partuclar case of Dicke Model.
The dynamics of entanglement for a system of two qubits or two-level atoms with collective Dicke states, has been widely studied in recent years \cite{He, Ficek08, Muzzamal2013, Yu}.  However, it is just a special kind and not the only kind of quantum correlation. Quantum discord (QD) is considered to be a more general resource of quantum information that captures the quantum or nonclassical correlations, even beyond entanglement, which may exists for mixed separable quantum states. A nonzero QD in some separable states has been considered as a useful resource in quantum state discrimination \cite{Li}, quantum locking \cite{Boixo} and is also responsible for the quantum computational efficiency of deterministic quantum computation with one pure qubit \cite{Lanyon}.
QD has been investigated in a wider context \cite{Streltsov, Yuan2010, Paterek,Ollivier2001,Henderson2001}, including quantum phase transitions in many-body physics \cite{Yuan2013, Wang}. Several other measurements have been introduced  in order to reveal the degree of correlation in quantum systems, such as local quantum uncertainty \cite{Girolami2013,Slaoui1,Slaoui2}, one-way quantum deficit \cite{Oppenheim2002,Horodecki2005}, local quantum Fisher information \cite{Kim2018}, quantum dissonance \cite{Modi2010}, measurement-induced nonlocality \cite{Luo2011}.
\par
The investigation of Bose-Einstein condensates (BEC) coupled to an optical cavity realizes the Dicke quantum phase transition in an open quantum system \cite{Baumann}. The creation of quantum discord by applying a Bell-like detection on optical cavities and the amplification of QD via cavity-BEC for initially coupled or uncoupled qubits has been proposed in Ref. \cite{Arani, Yuan2010, Yuan2013}.
The dynamics of a single impurity and the collective dephasing of a two spatial states impurity in BEC has been explored \cite{Mulansky, Cirone}.
The study of macroscopic nature of BEC with potential applications in quantum information is the dark- soliton (DS), resulting from the precise balance between the nonlinear and dispersive effects in the system \cite{Anderson, Kivshar}.
Recently, we introduced a dark soliton qubit in a quasi one dimensional (1D) BEC, being excellent candidates to store information given their appreciably long lifetimes ($\sim 0.01-1$s) \cite{Muzzamal2017}. The dark-soliton qutrit with unique properties of transmission and dispersion revealing the possibility of slowing down the speed of acoustic pulse has discussed in Ref. \cite{muzzamal2018a}. We explore the creation of entanglement between DS qubits by using the superposition of two maximally entangled states in the dissipative process of spontaneous emission \cite{muzzamal2018}. Moreover, the collective behaviour of the DS qubits reveals the dependence of entanglement evolution on the interatomic distance \cite{Muzzamal2018}.\par
In this paper, the quantum correlation behavior of two dark-soliton qubits is investigated with special attention to quantum discord. As described in Ref. \cite{Muzzamal2017}, the qubit is produced by trapping the impurities inside the potential created by dark-soliton, whereas the quantum fluctuations (phonons) play the role of a proper quantum reservoir. We show the generation and decay of correlations by computing the time evolution of quantum discord and classical correlations. We further demonstrate the dependence of correlations on the distance between dark-soliton qubits due to which Dicke symmetric and antisymmetric state decays to become subradiant and superradiant, respectively. \par
The paper is organised as follows: In Sec. II, the mean-field model of two DS qubits, placed in a quasi-1D BEC is discussed in detail, where we also compute the interaction between phonons and DS qubits. Sec. III describes the different measures of correlations, such as concurrence, quantum discord and classical correlations.
Sec. IV is devoted to the effect of collective or Dicke states on the time evolution of correlations. Discussion and conclusions are then provided in Sec. V.
\section{Theoretical Model.} We consider two DS qubits
immersed at distance $d$ in a quasi-1D BEC, surrounded by a dilute set of impurities and described by the wave functions $\Psi(x,t)$ and $\Phi(x,t)$, respectively \cite{muzzamal2018}. The impurities are considered to not interact, regarded as free particles and feel the soliton as potential \cite{Muzzamal2017}.
 At the mean-field level, the system is governed by the Gross-Pitaevskii and the Schr\"odinger equations, respectively
\begin{eqnarray}
i\hbar \frac{\partial \Psi}{\partial t}&=&-\frac{\hbar ^{2}}{2M}\frac{
\partial^{2} \Psi}{\partial x^{2}}+g\left\vert \Psi\right\vert ^{2}\Psi
+\chi\left\vert \Phi\right\vert ^{2}\Psi, \nonumber \\
i\hbar \frac{\partial \Phi}{\partial t}&=&-\frac{\hbar ^{2}}{2m}\frac{
\partial^{2} \Phi}{\partial x^{2}}+\chi\left\vert \psi _{\rm sol}\right\vert ^{2}\Phi.  \label{gp}
\end{eqnarray}
where $\chi$ and $g$ denotes the BEC-impurity coupling and BEC particle-particle interaction strength. $M$,$m$ represents the BEC particle and impurity masses, respectively and the singular nonlinear solution corresponds to the $i$th ($i=1,2$) soliton profile is  $\psi^{(i)}_{\rm sol}(x)=\sqrt{n_{0}}\tanh \left[ (x -x_i)/\xi \right]$ \cite{zakharov72, huang}, where $x_{i}=\pm d/2$ denotes the position of the respective soliton, $\xi =\hbar /\sqrt{Mn_{0}g}$ is the healing length and $n_{0}=10^8-10^9m^{-1}$ is the linear density in elongated $^{87}$Rb BECs. Here, the discussion is restricted to repulsive interactions ($g > 0$), where the dark solitons are assumed to be not disturbed by the presence of impurities. The impurity gas is choosed to be sufficiently dilute, i.e. $\vert\Phi\vert^2 \ll \vert\Psi\vert^2$ and much massive than the BEC particles. Such a situation can be produced, for example, choosing $^{134}$Cs impurities in a $^{87}$Rb BEC \cite{Michael2015}.
The experimentally accesible trap frequencies amount to be  $ \omega _{z}/2\pi= (15-730)$ Hz $\ll \omega_{r}/2\pi= (1-5)$ kHz and the corresponding length considered to be $l_{z}=(0.6-3.9)$ $\mu$m \cite{parker2004}. More recent experiments make eventual trap inhomogeneities to be much less critical by creating much larger traps, $l_z\sim 100$ $\mu$m \cite{schmiedmayer2010}. \par
For well separated solitons, i.e. $d \gg \xi$, the internal level structure of the two qubits is assumed to be equal. Each qubit is characterized by its ground state $g_i$ and excited levels $e_i$ separated by a gap frequency $\omega_0=\hbar \left(2\nu-1\right)/2m\xi^2$, where
\begin{eqnarray}
2\nu= -1+\sqrt{\frac{gM+4m\chi}{gM}}
\end{eqnarray}
is a parameter controlling the number of bound states $n_{bound}=1+\nu+\sqrt{\nu\left(\nu+1\right)}$, which operate as two-level system (qubit) in the range $0.33\leq \nu< 0.80$ \cite{Muzzamal2017}.
\\
{\bf Quantum Fluctuations:} The total condensate quantum field includes the DS field and quantum fluctuations, $\Psi(x)= \psi^{(i)}_{\rm sol}(x)+\delta \psi_{i}(x)$ with
\begin{eqnarray}
\delta \psi_{i}(x)=\sum_k \left(u^{(i)}_k(x) b_k +v^{*(i)}_k(x)b^{\dagger}_k \right)
\end{eqnarray}
and $b_k$'s denote the bosonic operators verifying the commutation relation $[b_{k},b^{\dagger}_{q}]=\delta_{k,q}$. The amplitudes $u_k(x)$ and $v_k(x)$ satisfy the normalization condition $\vert u_k(x)\vert ^2 -\vert v_k(x)\vert ^2=1$ and are explicitly given in \cite{Martinez2011}.
Therefore, the total Hamiltonian reads as the sum of qubit Hamiltonian $H_{\rm q}=\hbar \omega _{0}\left(\sigma _{z} ^{1}+ \sigma _{z} ^{2} \right)$ with the spin operator $\sigma_z=a_1^\dagger a_1- a_0^\dagger a_0$, the phonon (reservoir) Hamiltonian $H_{\rm p}=\sum_k \epsilon _{k}b_{k}^{\dagger}b_{k}$ with the Bogoliubov spectrum  $\epsilon _{k}=\mu \xi \sqrt{k^{2}(\xi^{2}k^{2}+2)}$, chemical potential $\mu=gn_{0}$ and the interaction Hamiltonian between DS and phonons given by \cite{Muzzamal2018},
 \begin{eqnarray*}
H_{\rm int}^{(1)} &=&\sum_{k}\sum_{i=1}^{2}\left( g_{l,l^{'}}^{(i)}(k)\sigma^{(i)} _{+}b_{k}+ g_{l,l^{'}}^{(i) \ast}(k) \sigma^{(i)} _{-}b_{k}^{\dag }\right)+{\rm h.c.}
\end{eqnarray*}
Here, $\sigma_{+}=\sigma_{-}^{\dagger}=a_{1}^{\dagger}a_{0}$ and
\begin{eqnarray*}
g_{l,l^{'}}^{(i)}(k) = \sqrt{n_{0}}\chi\int dx\varphi^{ \dag} _{l}(x) \varphi_{l^{'}}(x) \tanh\left(\frac{x -x_i}{\xi }\right) u^{(i)}_{k},
\end{eqnarray*}
with $\varphi_0(x)=A_0{\rm sech}  ^{\alpha}(x/\xi)$ and $\varphi_1(x)=2 A_1\tanh  (x /\xi) \varphi_0(x)$,
with $A_j (j=0,1)$ being the normalization constants, given by
\begin{eqnarray}
A_0&=&\left(\frac{\sqrt{\pi}  \Gamma[\alpha]}{\Gamma[\frac{1+2\alpha}{2}]}\right)^{-\frac{1}{2}}, \nonumber \\
A_1&=&\left(2^{2(1+\alpha)}A_0^2 \left(  \frac{ {_{2}}F_1[\alpha,2(1+\alpha),1+\alpha,-1]}{\alpha} \right.\right.  \nonumber \\   && \left. \left.
-\frac{ {_{2}}F_1[1+\alpha,2(1+\alpha),2+\alpha,-1]}{1+\alpha}\right.\right.  \nonumber \\   && \left. \left. +\frac{ {_{2}}F_1[2+\alpha,2(1+\alpha),3+\alpha,-1]}{2+\alpha}\right)\right)^{-\frac{1}{2}}, \label{normalization}
\end{eqnarray}
where $\Gamma[\alpha]$ and ${_{2}}F_1$ represents the Gamma and Hypergeometic function, respectively, and $\alpha=\sqrt{2g_{12}m_2/g_{11}m_1}$ determines the width of the wavefunction. The counter-rotating terms proportional to $b_{k}\sigma^{(i)} _{-}$ and $b^{\dagger}_{k}\sigma^{(i)} _{+}$ that do not conserve the total number of excitations are dropped by invoking the rotating wave approximation (RWA) which is  well justified in Ref. \cite{Muzzamal2017}.
\section{Quantifiers of quantum correlations}
Entanglement describes the non classical correlation between quantum systems. The study of entanglement and in particular how it can be quantified is a central topic within quantum information theory. Therefore, a most widely accepted measure for a two qubit system is the concurrence defined by Wootter's \cite{Wootters1998},
\begin{equation}
C\left( \rho\right)  = \max \left\{ {0,\sqrt{\vartheta _1} - \sum_{j=2}^4 \sqrt{\vartheta _j}} \right\}, \label{C}
\end{equation}
with ${\vartheta _i}$'s are the eigenvalues (in decreasing order) of the Hermition matrix $R=\rho \tilde \rho $, where the spin flip density matrix $\tilde \rho  = \left( {{\sigma _y} \otimes {\sigma _y}} \right){\rho ^*}\left( {{\sigma _y} \otimes {\sigma _y}} \right)$, with ${\rho ^*}$ and ${\sigma _y}$ being the complex conjugate of $\rho$ and the Pauli matrix, respectively. The concurrence is zero for separable states. The entanglement dynamics for two dark-soliton qubits has described in Ref.'s \cite{muzzamal2018, Muzzamal2018}.
\par
Non-classical correlations can exist even in separable mixed states for which the Wootters Concurrence is zero. Thus, to go beyond Wootters formula (\ref{C}), the quantum discord was introduced in \cite{Ollivier2001,Henderson2001} as the difference betwen the quantum mutual information and classical correlations. The quantum mutual information is the quantum analog of the classical mutual information defined by,
\begin{equation}
I(A,B)  = H(A)+H(B)-H(A,B), \label{classical information}
\end{equation}
where $A$, $B$ denotes the random variables with given probability distribution and $H=-\sum_n p_n {\rm log}p_n$ describes the Shannon entropy. To generalize this concept to quantum physics, the quantum mutual information for bipartite quantum systems is defined by,
\begin{equation}
\mathcal{I}(\rho_{AB})  = S(\rho_A)+S(\rho_B)-S(\rho_{AB}), \label{classical information}
\end{equation}
where the von Neumann entropy $S(\rho_A)=-{\rm tr} \rho_A {\rm log}\rho_A$ replaces the Shannon entropy in the classical variant of mutual information. Quantum mutual information may also be written as the sum of quantum ($\mathcal{Q}\left( {{\rho _{AB}}} \right)$) and classical ($\mathcal{C}\left( {{\rho _{AB}}} \right)$) correlations
\begin{equation}
\mathcal{Q}\left( {{\rho _{AB}}} \right): = \mathcal{I}\left( {{\rho _{AB}}} \right) - \mathcal{C}\left( {{\rho _{AB}}} \right),
\end{equation}
Here, $\mathcal{Q}\left( {{\rho _{AB}}} \right)$ is called quantum discord and the quantity $\mathcal{C}\left( {{\rho _{AB}}} \right)$ is defined as
\begin{equation}
\mathcal{C}\left( {{\rho _{AB}}} \right)=\mathop {\max }\limits_{{\pi _B}^j} \left[ {S\left( {{\rho _A}} \right) - \sum\limits_j {{p_j}S\left( {\rho _A^j} \right)} } \right], \label{cla}
\end{equation}
where
${{\pi _B}^j}$ is a set of local projective measurements on the subsystem $B$, ${p_{B,j}} = t{r_{AB}}\left[ {{\rho _{AB}}\left( {I \otimes {\pi _B}^j} \right)} \right]$ and ${\rho _{B,j}} = t{r_A}\left[ {\left( {I \otimes {\pi _B}^j} \right){\rho _{AB}}\left( {I \otimes {\pi _B}^j} \right)} \right]$ are the probability and the conditional state of system $B$ associated with outcome $j$. Quantum discord is an efficient measure of nonclassical correlations that may include entanglement but is not always larger than entanglement \cite{Li2007, Luo}. This indicates that discord is not simply a sum of entanglement and some other nonclassical correlation. Even for the simplest case of two entangled qubits, the relation between quantum discord, entanglement, and classical correlation is not yet clear.
The entropic quantum discord of any 2-qubit rank states can be calculated exactly. However, the situation becomes complicated for states with rank large than two because of the difficulty to minimize the conditional entropy of subsystem $A$ (or subsystem $B$). \par

Recently, the linear entropy approximation (Renyi-2 entropy) was employed to find the classical correlations of qudit-qubit states \cite{Ma2015}. It has been found that the quantum discord based on linear entropy has deep connection with the original discord defined by von Neumann entropy in which average deviation is of the order $10^{-4}$.
Here, we study the classical correlations based on the linear entropy
with which we shall derive the analytical expression of the quantum discord for any arbitrary two-qubit states.
The linear entropy of a state $\rho$ is given by
\begin{equation}
    {S_2}\left( \rho  \right) = 2\left[ {1 - {\rm tr}\left( {{\rho ^2}} \right)} \right]. \label{entropy}
\end{equation}
To compute the quantum discord, we first write a bipartite quantum state $\rho _{AB}$ as  \cite{Osborne2006}
\begin{equation}
{\rho _{AB}} = {\Lambda _\rho } \otimes {I_B}\left( {\left| {{r_{B'B}}} \right\rangle \left\langle {{r_{B'B}}} \right|} \right),
\end{equation}
where ${\Lambda _\rho }$ is a qudit channel, maps a qubit state $B'$ to the qudit state $A$ and ${\left| {{r_{B'B}}} \right\rangle }$ represents the symmetric two-qubit purification of the reduced density operator ${\rho _B}$, i.e., ${r_{B'}} = {r_B} = {\rho _B}$. In the spectral decomposition,
\begin{equation}
{\rho _B} = \sum\limits_{i = 0,1} {{\lambda _i}\left| {{\varphi _i}} \right\rangle \left\langle {{\varphi _i}} \right|} \label{rho_B}
\end{equation}
 with $\left| {{\varphi _0}} \right\rangle  = {\left( {{a_0},{a_1}} \right)^T}$ and $\left| {{\varphi _1}} \right\rangle  = {\left( {{b_0},{b_1}} \right)^T}$
 , where $T$ stands for transpose. Generally, a qudit state in Bloch expression is given by $\rho  = {{\left( {{I_d} + \vec r\gamma } \right)} \mathord{\left/
        {\vphantom {{\left( {{I_d} + \vec r\gamma } \right)} d}} \right.
        \kern-\nulldelimiterspace} d}$, where $I_d$ denotes the $d\times d$ identity matrix, $\vec r$ is a $d^2-1$ dimensional real vector, and $\gamma  = {\left( {{\gamma _1},{\gamma _2},...,{\gamma _{{d^2} - 1}}} \right)^T}$ is the vector of the generators of  lie algebra $SU(d)$. Using Eq. (\ref{entropy}), it's easy to show that
\begin{equation}
    {S_2}\left( {\frac{{{I_d} + \vec r\gamma }}{d}} \right) = \frac{{2{d^2} - 2d - 4{{\left| {\vec r} \right|}^2}}}{{{d^2}}}.
\end{equation}
Similarly, a qubit state can be written as
$\rho  = {{\left( {{I_2} + {{\vec r}_B}\sigma } \right)} \mathord{\left/
        {\vphantom {{\left( {{I_d} + {{\vec r}_B}\sigma } \right)} 2}} \right.
        \kern-\nulldelimiterspace} 2}$, where $\sigma  = {\left( {{\sigma _1},{\sigma _2},{\sigma _3}} \right)^T}$ denotes the Pauli operators. The qubit channel ${\Lambda _\rho } = {{\left( {{I_d} + \left( {L{{\vec r}_B} + l} \right)\gamma } \right)} \mathord{\left/
{\vphantom {{\left( {{I_d} + \left( {L{{\vec r}_B} + l} \right)\gamma } \right)} d}} \right. \kern-\nulldelimiterspace} d}$ with the three dimensional $l$ vector and $(d^2-1)\times3$ real matrix with elements  ${L_{ij}} = {{{\rm tr}\left[ {\Lambda \left( {{\sigma _j}} \right){\sigma _i}} \right]} \mathord{\left/
{\vphantom {{tr\left[ {\Lambda \left( {{\sigma _j}} \right){\sigma _i}} \right]} 2}} \right.
\kern-\nulldelimiterspace} 2}$. Therefore, it is simple to see that the reduced density matrix of the qudit state $\rho_A$ is exactly equal to the completely positive trace-preserving map of the reduced density matrix $\rho_B$, i.e., ${\rho _A} = \Lambda \left( {{\rho _B}} \right)$. Moreover, Eq. (\ref{cla}) coincides with the linear Holevo capacity for qubit channel ${\chi _2}\left( {{\rho _B},\Lambda } \right)$ (see Eq.(4) of Ref. \cite{Osborne2006}). Based on these results, the analytical expression of the classical correlation of arbitrary $d\otimes 2$ quantum states is reduced to
\begin{equation}
    {{\cal C}_2}\left( {{\rho _{AB}}} \right) = \frac{4}{d}{\lambda _{\max }}\left( {{L^T}L} \right){S_2}\left( {{\rho _B}} \right), \label{cla2}
\end{equation}
where ${\lambda _{\max }}\left( {{L^T}L}\right)$ stands for the largest eigenvalue of the matrix ${L^T}L$. Now, we give the main lines to compute the elements of the matrix $L$ occurring in (\ref{cla2}) for an arbitrary qubit-qubit system. In the computational basis $\left\{ {\left| {00} \right\rangle ,\left| {01} \right\rangle ,\left| {10} \right\rangle ,\left| {11} \right\rangle } \right\}$, the density matrix of two qubit states can be written as
\begin{equation}
    {\rho _{AB}} = {\beta _{00}} \otimes \left| 0 \right\rangle \left\langle 0 \right| + {\beta _{01}} \otimes \left| 0 \right\rangle \left\langle 1 \right| + {\beta _{10}} \otimes \left| 1 \right\rangle \left\langle 0 \right| + {\beta _{11}} \otimes \left| 1 \right\rangle \left\langle 1 \right|,   \label{beta11}
\end{equation}
with
\begin{eqnarray}
{\beta _{00}} = \left( {\begin{array}{*{20}{c}}
        {{\rho _{11}}}&{{\rho _{13}}}\\
        {{\rho _{31}}}&{{\rho _{33}}}
        \end{array}} \right), \hspace{0.5cm}  {\beta _{01}} = \left( {\begin{array}{*{20}{c}}
{{\rho _{21}}}&{{\rho _{23}}}\\
{{\rho _{41}}}&{{\rho _{43}}}
\end{array}} \right), \nonumber \\
{\beta _{10}} = \left( {\begin{array}{*{20}{c}}
        {{\rho _{12}}}&{{\rho _{14}}}\\
        {{\rho _{32}}}&{{\rho _{34}}}
        \end{array}} \right), \hspace{0.5cm} {\beta _{11}} = \left( {\begin{array}{*{20}{c}}
{{\rho _{22}}}&{{\rho _{24}}}\\
{{\rho _{42}}}&{{\rho _{44}}}
\end{array}} \right).
\end{eqnarray}
According to the symmetric two-qubit purification of the reduced density operator $\rho _B$ on an auxiliary qubit system $B'$, one  can determine $\left| {{r_{B'B}}} \right\rangle$ from Eq. (\ref{rho_B}),
\begin{equation}
\left| {{r_{B'B}}} \right\rangle \left\langle {{r_{B'B}}} \right| = \left( {\begin{array}{*{20}{c}}
        {A\bar A}&{AB}&{A\bar B}&{AD}\\
        {B\bar A}&{B\bar B}&{B\bar B}&{B\bar D}\\
        {B\bar A}&{B\bar B}&{B\bar B}&{B\bar D}\\
        {D\bar A}&{D\bar B}&{D\bar B}&{D\bar D}
        \end{array}} \right),
        \label{vector}
\end{equation}
with
\begin{equation}
\left\{ \begin{array}{l}
A = \sqrt {{\lambda _1}} {a_0}^2 + \sqrt {{\lambda _2}} {b_0}^2,\\
B = \sqrt {{\lambda _1}} {a_0}{a_1} + \sqrt {{\lambda _2}} {b_0}{b_1},\\
D = \sqrt {{\lambda _1}} {a_1}^2 + \sqrt {{\lambda _2}} {b_1}^2.
\end{array} \right.
\end{equation}
Therefore, the elements of Eq. (\ref{beta11})  can be manipulated with Eq. (\ref{vector}) as
\begin{eqnarray}
    {\beta _{00}} &=& A\bar A\Lambda \left( {\left| 0 \right\rangle \left\langle 0 \right|} \right) + A\bar B\Lambda \left( {\left| 0 \right\rangle \left\langle 1 \right|} \right) \nonumber \\
    &+&
     B\bar A\Lambda \left( {\left| 1 \right\rangle \left\langle 0 \right|} \right) + B\bar B\Lambda \left( {\left| 1 \right\rangle \left\langle 1 \right|} \right), \nonumber \\
     {\beta _{01}} &=& B\bar A \Lambda\left( {\left| 0 \right\rangle \left\langle 0 \right|} \right) + B\bar B \Lambda \left( {\left| 0 \right\rangle \left\langle 1 \right|} \right)  \nonumber \\
     &+& D\bar A \Lambda \left( {\left| 1 \right\rangle \left\langle 0 \right|} \right) + D\bar B \Lambda \left( {\left| 1 \right\rangle \left\langle 1 \right|} \right), \nonumber \\
     {\beta _{10}} &= &A\bar B\Lambda \left( {\left| 0 \right\rangle \left\langle 0 \right|} \right) + A\bar D\Lambda \left( {\left| 0 \right\rangle \left\langle 1 \right|} \right) \nonumber \\
     &+& B\bar B\Lambda \left( {\left| 1 \right\rangle \left\langle 0 \right|} \right) + B\bar D\Lambda \left( {\left| 1 \right\rangle \left\langle 1 \right|} \right),\nonumber \\
     {\beta _{11}} &= &B\bar B\Lambda \left( {\left| 0 \right\rangle \left\langle 0 \right|} \right) + B\bar D\Lambda \left( {\left| 0 \right\rangle \left\langle 1 \right|} \right) \nonumber \\
     &+& D\bar B\Lambda \left( {\left| 1 \right\rangle \left\langle 0 \right|} \right) + D\bar D\Lambda \left( {\left| 1 \right\rangle \left\langle 1 \right|} \right).\label{beta}
\end{eqnarray}
The elements of the matrix $L$ can be determined by solving Eq. (\ref{beta}) to get $\Lambda \left( {\left| i \right\rangle \left\langle j \right|} \right)$ from which we obtains the elements ${L_{ij}} = {{{\rm tr}\left[ {\Lambda \left( {{\sigma _j}} \right){\sigma _i}} \right]} \mathord{\left/
		{\vphantom {{tr\left[ {\Lambda \left( {{\sigma _j}} \right){\sigma _i}} \right]} 2}} \right.
		\kern-\nulldelimiterspace} 2}$ in terms of the density matrix elements.



\section{Measurement of correlations with Dicke States}
We consider here the most adequate so called collective Dicke states \cite{Dicke1954} described by $ \left| e \right\rangle=\left| {{e_1},{e_2}} \right\rangle, \left| s \right\rangle=\left( \left| {{e_1},{g_2}} \right\rangle + \left| {{g_1},{e_2}} \right\rangle \right)/\sqrt{2}, \left| a \right\rangle=\left( \left| {{e_1},{g_2}} \right\rangle - \left| {{g_1},{e_2}} \right\rangle \right)/\sqrt{2}$ and $\left| g \right\rangle=\left| {{g_1},{g_2}} \right\rangle$,
to measure the time evolution of correlations. The density matrix corresponds to these collective states can be written as
 \begin{equation}
\rho_{q}(t) =\left(
\begin{array}{cccc}
\rho _{ee} & \rho _{es} & \rho _{ea} & \rho _{eg} \\
\rho _{se} & \rho _{ss} & \rho _{sa} & \rho _{sg} \\
\rho _{ae} & \rho _{as} & \rho _{aa} & \rho _{ag} \\
\rho _{ge} & \rho _{gs} & \rho _{ga} & \rho _{gg}
\end{array}%
\right) ,  \label{Dens. Mat Dicke}
\end{equation}
with $\rho_{ij} =\left \langle \psi_{i} \right\vert \rho \left\vert \psi_{j} \right \rangle $, where $i,j=e,s,a,g$. We derive the master equation \cite{Muzzamal2018} to describe the density matrix elements after taking trace over the phonons degrees of freedom,
\begin{eqnarray}
\frac{\partial\rho_{q}(t)}{\partial t}  &=& -\frac{i}{\hbar} \left[H_{q},\rho_{q}(t)\right] -\sum^{2}_{i\neq j}\eta_{ij}\left[\sigma_{+}^{i}\sigma_{-}^{j},\rho_{q}(t)\right]\nonumber \\&+&  \sum^{2}_{ij=1}\Gamma_{ij}\left[\sigma_{-}^{j}\rho_{q}(t)\sigma_{+}^{i}-\frac{1}{2} \lbrace \sigma_{+}^{i}\sigma_{-}^{j},\rho_{q}(t) \rbrace \right],
 \label{master eq.}
\end{eqnarray}
where
\begin{eqnarray}
\Gamma_{ij} &=& 2L\int_{0}^{\infty}dk g^{(i)}_{k}g_{k}^{(j)*} \delta(\omega_{k}-\omega_{0}) \nonumber \\
\eta_{ij} &=&\frac{L}{2\pi}\wp\int_{0}^{\infty}dk g^{(i)}_{k}g_{k}^{(j)*}\frac{1}{\left(\omega _{k}-\omega _{0}\right)},\label{parameters}
\end{eqnarray}
with $\wp$ standing for the principal value of the integral. The term $\Gamma_{ij}(=\gamma)$ is the damping of the $i$th atom by spontaneous emission for $i=j$ while $\Gamma_{ij}(=\Gamma)$ for $i \neq j$ depicts the collective damping resulting from the mutual exchange of spontaneously emitted phonon. The coherent term $\eta_{ij}(=\eta)$ represents the phonon-induced coupling between the qubits.
\begin{figure}[t!]
\includegraphics[scale=0.49]{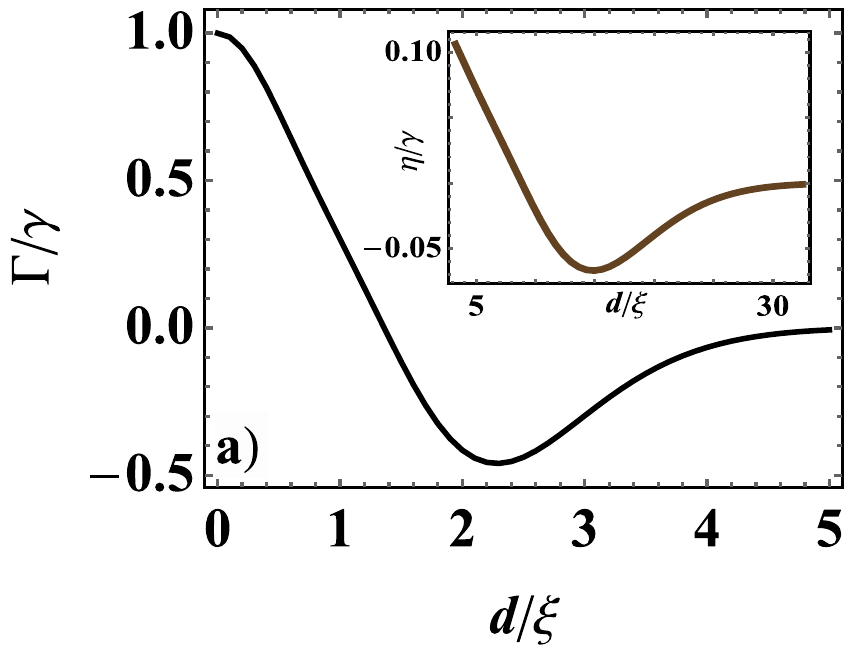}
\includegraphics[scale=0.49]{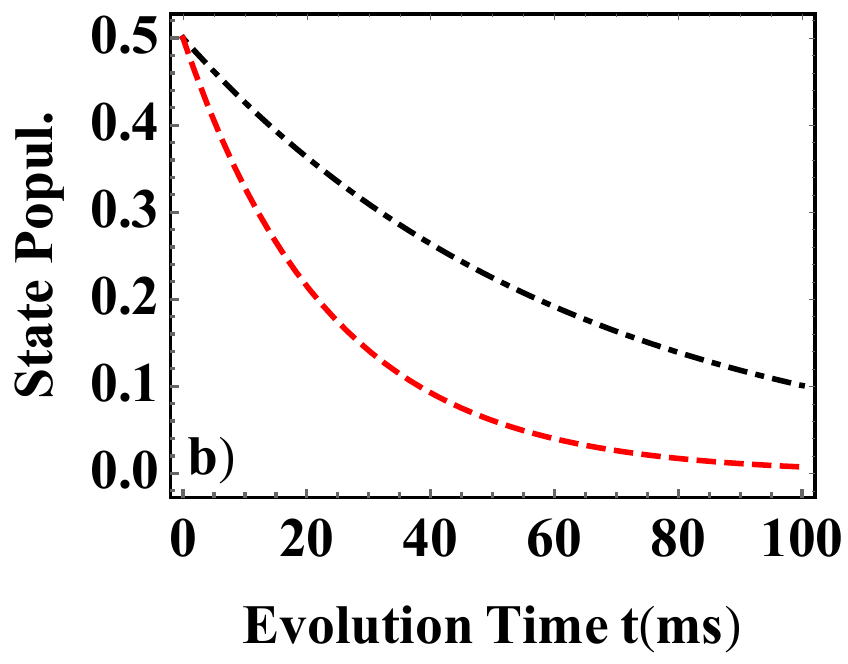}
\caption{(color online) Panel a) shows the collective damping $\Gamma$ and qubit-qubit interaction $\eta$ (inset) as a function of interatomic distance $d$. Panel b) depicts the population of antisymmetric  state $\left\vert a\right\rangle$ (dashed curve) and symmetric state $\left\vert s\right\rangle$ (dotted-dashed curve) at distance $d \simeq 5\xi/2$}
\label{fig_Gamma}
\end{figure}
Therefore, the density matrix elements by using Dicke states can be written as
\begin{eqnarray}
\rho_{ee}(t)&=& e^{-2\gamma t}\rho_{ee}(0)\nonumber \\
\rho_{ss}(t)&=&e^{-\left(\gamma+\Gamma\right) t}\rho_{ss}(0) \nonumber \\ &+& \frac{\left(\gamma+\Gamma\right)}{\left(\gamma-\Gamma\right)}\left(e^{-\left(\gamma+\Gamma\right) t}-e^{-2\gamma t}\right)\rho_{ee}(0)\nonumber \\
\rho_{aa}(t)&=&e^{-\left(\gamma-\Gamma\right) t}\rho_{aa}(0)\nonumber \\ &+& \frac{\left(\gamma-\Gamma\right)}{\left(\gamma+\Gamma\right)}\left(e^{-\left(\gamma-\Gamma\right) t}-e^{-2\gamma t}\right)\rho_{ee}(0)\nonumber \\
\rho_{eg}(t)&=& e^{-\gamma t}\rho_{eg}(0)\nonumber \\
\rho_{sa}(t)&=& e^{-\left(\gamma-2i\eta\right) t}\rho_{sa}(0)
\label{den. mat ele.}
\end{eqnarray}
with $\rho_{gg}=1-\rho_{ee}-\rho_{ss}-\rho_{aa}$ and $\rho_{jk}=\rho^{*}_{kj}$. Eq. (\ref{den. mat ele.}) depicts that all transition rates to and from
the state $\rho_{ss}$ are equal to $\gamma+\Gamma$ (superradiant) while from state $\rho_{aa}$ (subradiant) are $\gamma-\Gamma$. In case of DSs, the state $\rho_{aa}$ decays with an enhanced
(superradiant) rate and $\rho_{ss}$ with a reduced (subradiant) rate at sufficiently large distance (see Fig. \ref{fig_Gamma}).
Here, the effect of promising function $\Gamma$ has analyzed to measure the correlation between two dark-soliton qubits for different initial states.
\subsection{Superposition of maximally entangled states}
As a first example, the system is prepared in the superposition of maximally entangled symmetric and anti-symmetric state, i.e.,
\begin{equation}
\left| {\psi } \right\rangle  = \frac{1}{{\sqrt 2 }}\left( {\left| s \right\rangle  + \left| a \right\rangle } \right),
\end{equation}
from which it decays spontaneously.
\begin{figure}[t!]
\includegraphics[scale=0.7]{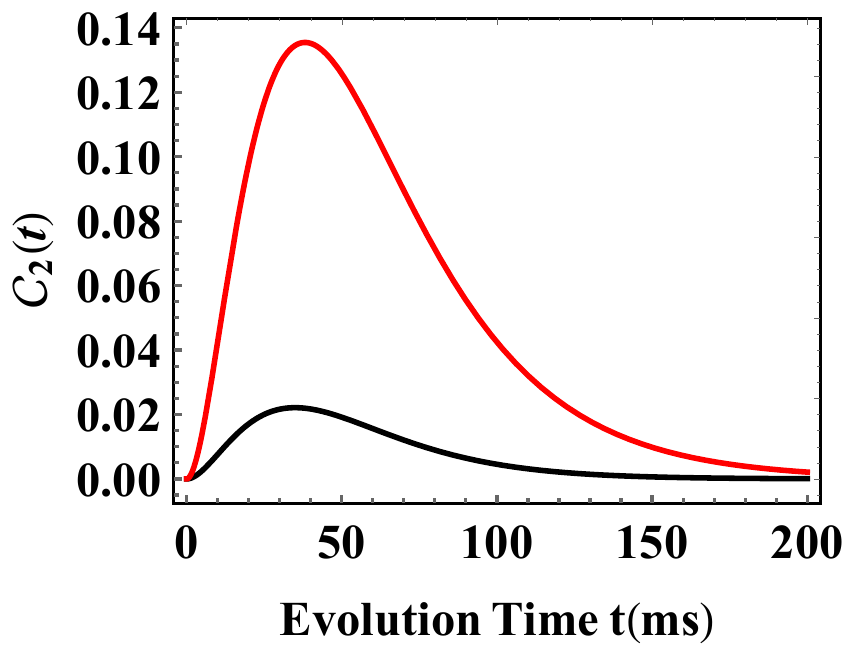}
\caption{(color online) Time evolution of the classical correlations $\mathcal{C}_2(t)$ at distance $d \sim 6\xi/5$ (black curve) and $d \sim 5\xi/2$ (red curve).}
\label{fig_cc}
\end{figure}
\begin{figure}[t!]
\includegraphics[scale=0.7]{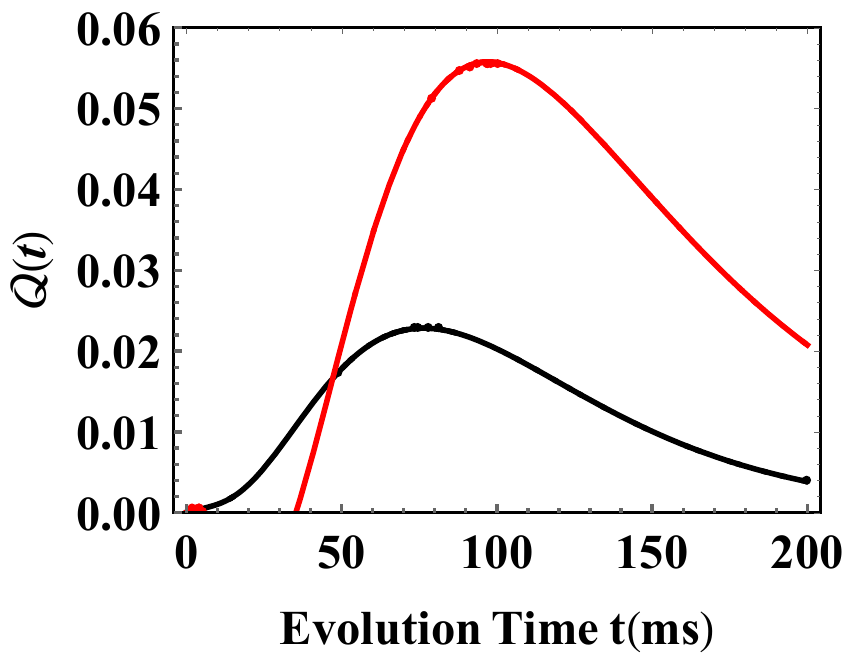}
\caption{(color online) Time evolution of the quantum discord $\mathcal{Q}(t)$ at distance $d \sim 6\xi/5$ (black curve) and $d \sim 5\xi/2$ (red curve).}
\label{fig_qd}
\end{figure}
\begin{figure}[t!]
\includegraphics[scale=0.7]{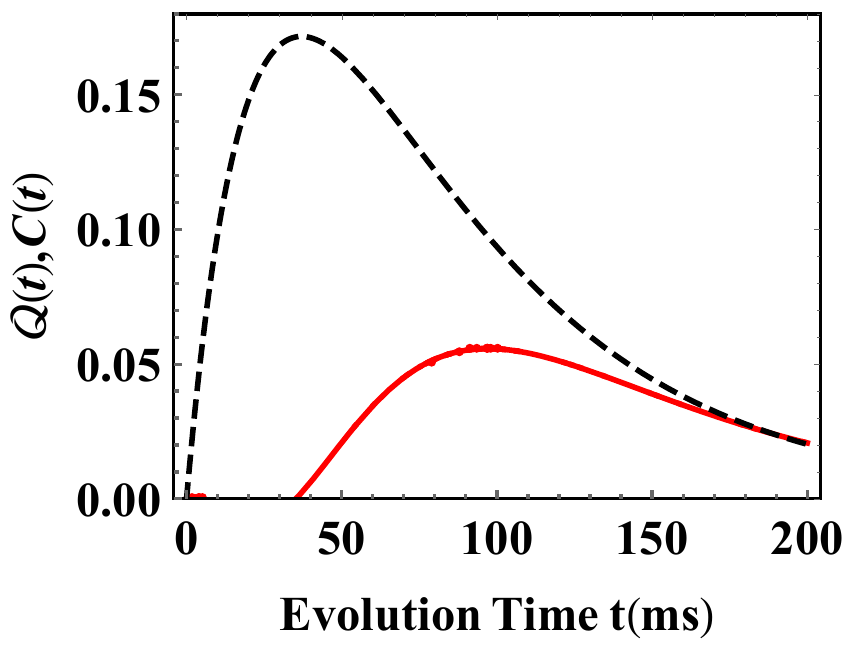}
\caption{(color online) Comparison of quantum discord $\mathcal{Q}(t)$ (solid curve) and concurrence $C(t)$ (dashed curve) at distance $d \sim 5\xi/2$.}
\label{fig_comp}
\end{figure}
To determine the analytic expression of classical correlation,
direct computation shows $L = {\rm diag}\left\{ {{L_{11}},{L_{22}},{L_{33}}} \right\}$ with
\begin{eqnarray}
{L_{11}} &=& \frac{{ - 2sinh\left( {\Gamma t} \right)}}{{\sqrt {\kappa_- \left( {2{e^{\gamma t}}  -\kappa_- } \right)} }} , \nonumber \\
{L_{22}} &=& \frac{{2isin\left( {2\eta t} \right)}}{{\sqrt {\kappa_-\left( {2{e^{\gamma t}} -\kappa_-} \right)} }}, \nonumber \\
{L_{33}} &=& \frac{{\kappa_+}}{{\kappa_-- 2{e^{\gamma t}} }},
\end{eqnarray}
where $\kappa_{\pm}=\cosh \left( {\Gamma t}\right) \pm \cos \left( {2\eta t} \right)$.
Therefore, the classical correlations under the linear entropy takes the form
\begin{align}
{{\cal C}}_2\left( {t} \right) &={S_2}\left( {{\rho _B}} \right) \max\left[ {{L_{11}^2},{L_{22}^2},{L_{33}^2}} \right],  \label{superposition cc}
\end{align}
where ${S_2}\left( {{\rho _B}} \right) = {\kappa_- e^{ - 2\gamma t}}\left( {2{e^{\gamma t}} + \kappa_- } \right)$. Thus, the analytical expression of the quantum discord is given by
\begin{align}
\mathcal{Q}\left( {t} \right) &=  -\sum_{i=\pm} {\beta _ {\pm} }{\log _2}\left( {{\beta _ {\pm} }} \right)  +\sum\limits_{j = 1}^4 {{\zeta _j}} {\log _2}\left( {{\zeta _j}} \right) \notag\\&- \sum_{l=\pm} {\xi _ {\pm} }{\log _2}\left( {{\xi _ {\pm} }} \right) -{{\cal C}_2}\left( {{t}} \right), \label{superposition qd}
\end{align}
with
\begin{eqnarray}
{\beta _ \pm } &=& \frac{{{e^{ - \gamma t}}}}{2}\kappa_{\pm}, \nonumber \\
{\xi _ \pm } &=& \frac{{{e^{ - \gamma t}}}}{2}\left( {2{e^{\gamma t}}- \kappa_{\mp}} \right),
\end{eqnarray}
and the eigenvalues ${\zeta _i}$ of the density matrix $\rho$ are given by
\begin{eqnarray}
{\zeta _1} &=&0, \hspace{2.8cm}  {\zeta _2} =1 - {e^{ - \gamma t}}\cosh \left( {\Gamma t} \right), \nonumber \\
{\zeta _{3,4}} &=& \frac{{{e^{ - \gamma t}}}}{2}\left( {\cosh \left( {\Gamma t} \right) \pm \sqrt {\cos \left( {4\eta t} \right) + \sinh {{\left( {\Gamma t} \right)}^2}} } \right). \nonumber \\
\end{eqnarray}
Fig. \ref{fig_qd} depicts the generation of  quantum correlation in terms of quantum discord by measuring the classical correlations of Fig. \ref{fig_cc}. It is observed that the quantum discord appears at  $\sim 40$ms, where the concurrence starts to decay and both reached to the same value of correlation at ($\sim 150$ms) large distance $d \sim5\xi/2 \sim2-5 \mu$m (see Fig. \ref{fig_comp}) for a BEC in the conditions of \citep{zoran2013}. It is pertinent to mention here that a major limitation to the quantum discord performance could be the DS quantum diffusion, or quantum evaporation \cite{Dziarmaga2004}, a feature that has been theoretically predicted but yet not experimentally validated.
\subsection{Entangled State}
In the second case, we choose the system to be in an entangled state, corresponds to a linear superposition of both or neither of the excited states, i.e.,
\begin{equation}
\left| \psi  \right\rangle  = \sqrt \alpha  \left| e \right\rangle  + \sqrt {1 - \alpha } \left| g \right\rangle     \label{entangled state}
\end{equation}
where $0 \le \alpha  \le 1$.
\begin{figure}[t!]
\includegraphics[scale=0.65]{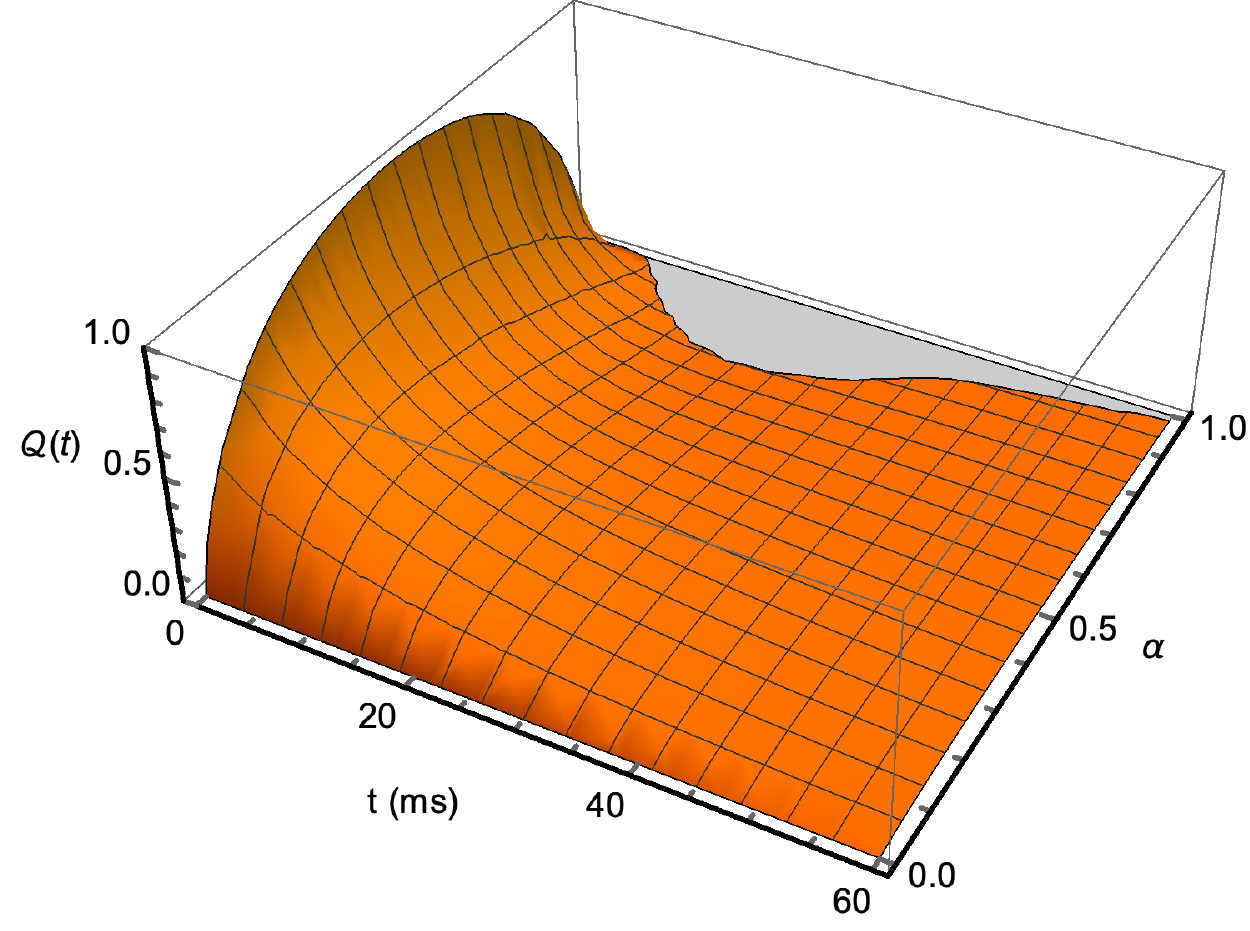}
\caption{(color online) Time evolution of quantum discord $\mathcal{Q}$(t) for non-maximally entangled state at distance $d \sim 5\xi/2$.}
\label{fig_Entangled}
\end{figure}
Similar to the first case, the classical correlation and quantum discord can be determined by using Eq. (\ref{superposition cc}) with the linear entropy $S_2(\rho_B)$, the elements of the matrix $L$  and the eigenvalues ${\zeta _{i}}$  of the density matrix ${{\rho _{AB}}}$, described in Appendix-\ref{Entangled State Elements}.

Fig. \ref{fig_Entangled} determines the time evolution of quantum discord for an initial non-maximally entangled state. The quantum discord follows asymptotic decay at large distance
 $d \gg \xi$ ($\Gamma=0$). The phenomenon of sudden death and revival for quantum discord occurs at $\alpha \geq 0.80$. By estimating the death and revival time with the use of few kHz of chemical potential $\mu$, the dark period (revival-sudden death) appears to be $10$ms which can be averted or delayed by carrying out some local unitary operators on DS qubits \citep{ Singh2017, Rau2008}.

\subsection{Mixed State}
Here, we describe the evolution of correlation by assuming a two-qubit system to be initially prepared in a diagonal basis of the collective states, i.e.,
\begin{equation}
\rho \left( 0 \right) = \frac{1}{3}\left( {\begin{array}{*{20}{c}}
    \alpha &0&0&0\\
    0&2&0&0\\
    0&0&0&0\\
    0&0&0&{1 - \alpha }
    \end{array}} \right).
\end{equation}
\begin{figure}[t!]
\includegraphics[scale=0.65]{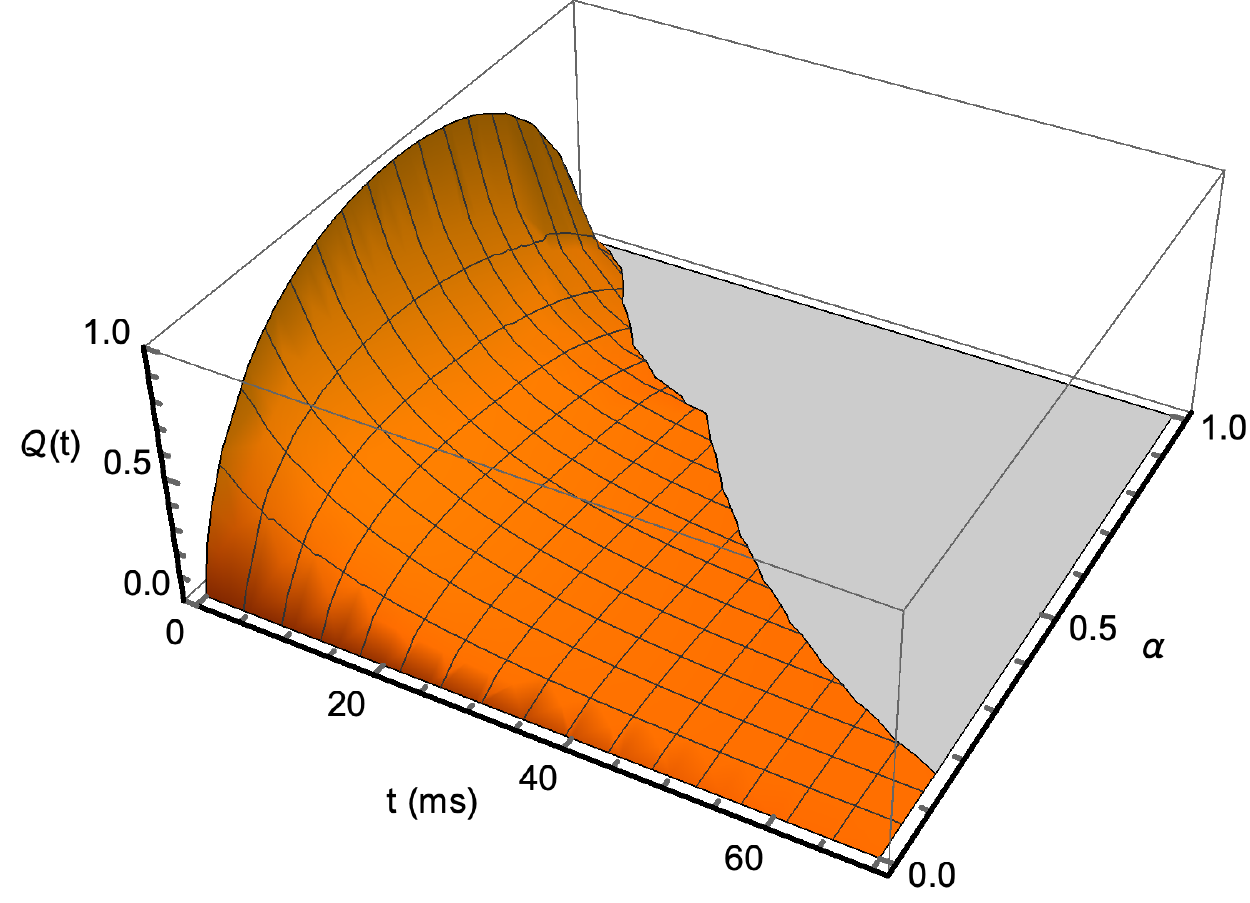}
\caption{(color online) Time evolution of quantum discord $\mathcal{Q}$(t) for a mixed initial state at distance $d \sim 5\xi/2$.}
\label{fig_Mixed}
\end{figure}
The derivation of the classical correlation  elements and quantum discord can be found in Appendix-\ref{Mixed State Elements}. Fig. \ref{fig_Mixed} shows the time evolution of quantum discord for a mixed initial state. It can be seen that the sudden death for quantum discord is possible only for $\alpha \geq 0.2$.

\section{Summary and Discussion}
To summarize, we have investigated the dynamics of quantum correlations in terms of quantum discord, by placing two dark-soliton qubits in a quasi one-dimensional Bose-Einstein condensates, surrounded by a dilute set of impurities. We derive the master equation and extracted the density matrix elements by using Dicke bases. The linear entropy approximation (Renyi-2 entropy) is used to find the analytical expression of the quantum discord for different initial states.  We have shown the spontaneous generation and decay of quantum correlation at sufficiently large distance $\sim \mu$m. This is the first study on the quantum discord, involving phononic degrees of freedom in dark-soliton qubit and we believe that the suggested approach opens a promising research avenue in the field of phononic information processing.

\appendix
\section{Entangled State Elements}
\label{Entangled State Elements}
The linear entropy $S_2(\rho_B)$ for an entangled state of Eq. (\ref{entangled state}) is given by
\begin{widetext}
\begin{eqnarray}
S_2\left( \rho _B \right) = \frac{2 e^{ - 2\gamma t}}{\left( \gamma ^2 - \Gamma ^2 \right)^2}
\left[
2\alpha \left( \gamma ^2 - \Gamma ^2 \right)^2\{1-\delta e^{  \gamma t}+2\alpha \delta e^{  -\gamma t}  \}  -2\alpha^2 \{\delta^2+\left( \gamma ^2 - \Gamma ^2 \right)^2 e^{  -2\gamma t}  \}  \right].
\end{eqnarray}
\end{widetext}
The elements of the matrix $L$  after simplification can be written as,
\begin{widetext}
  \begin{eqnarray}
{L_{11}} &=& \frac{{2\alpha \gamma \Gamma Z - \alpha \left( {{\gamma ^2} + {\Gamma ^2}} \right)\sinh \left( {\Gamma t} \right) + \left( {{\gamma ^2} - {\Gamma ^2}} \right)\sqrt {\alpha \left( {1 - \alpha } \right)} }}{{\sqrt {\alpha \left( {\left( {{\gamma ^2} - {\Gamma ^2}} \right)\left( {{e^{\gamma t}} - \alpha {e^{ - \gamma t}}} \right) - \alpha \delta } \right)\left( {\delta  + \left( {{\gamma ^2} - {\Gamma ^2}} \right){e^{ - \gamma t}}} \right)} }}, \nonumber \\
{L_{22}} &=& \frac{{2\alpha \gamma \Gamma Z - \alpha \left( {{\gamma ^2} + {\Gamma ^2}} \right)\sinh \left( {\Gamma t} \right) - \left( {{\gamma ^2} - {\Gamma ^2}} \right)\sqrt {\alpha \left( {1 - \alpha } \right)} }}{{\sqrt {\alpha \left( {\left( {{\gamma ^2} - {\Gamma ^2}} \right)\left( {{e^{\gamma t}} - \alpha {e^{ - \gamma t}}} \right) - \alpha \delta } \right)\left( {\delta  + \left( {{\gamma ^2} - {\Gamma ^2}} \right){e^{ - \gamma t}}} \right)} }},  \nonumber \\
{L_{33}} &=& \frac{{{{\left( {{\gamma ^2} - {\Gamma ^2}} \right)}^2}\left( {1 - \alpha {e^{ - 2\gamma t}}} \right) - \left( {{\gamma ^2} - {\Gamma ^2}} \right)2\alpha \delta {e^{ - 2\gamma t}} - \alpha {\delta ^2}}}{{\left( {\left( {{\gamma ^2} - {\Gamma ^2}} \right)\left( {{e^{\gamma t}} - \alpha {e^{ - \gamma t}}} \right) - \alpha \delta } \right)\left( {\delta  + \left( {{\gamma ^2} - {\Gamma ^2}} \right){e^{ - \gamma t}}} \right)}}.  \nonumber \\
 \end{eqnarray}
 \end{widetext}
In this case, the quantum discord takes the form
\begin{widetext}
\begin{align}
\mathcal{Q}\left( {{t}} \right) =& \frac{{ - 2\alpha {e^{ - \gamma t}}}}{{\left( {{\gamma ^2} - {\Gamma ^2}} \right)}}\left( {\delta  + \left( {{\gamma ^2} - {\Gamma ^2}} \right){e^{ - \gamma t}}} \right){\log _2}\left( {\frac{{\alpha {e^{ - \gamma t}}}}{{\left( {{\gamma ^2} - {\Gamma ^2}} \right)}}\left( {\delta  + \left( {{\gamma ^2} - {\Gamma ^2}} \right){e^{ - \gamma t}}} \right)} \right) + \sum\limits_{i = 1}^4 {{\zeta _i}} {\log _2}\left( {{\zeta _i}} \right) -\notag\\& \frac{{2{e^{ - \gamma t}}}}{{\left( {{\gamma ^2} - {\Gamma ^2}} \right)}}\left( {\left( {{\gamma ^2} - {\Gamma ^2}} \right)\left( {{e^{\gamma t}} - \alpha {e^{ - \gamma t}}} \right) - \alpha \delta } \right){\log _2}\left( {\frac{{{e^{ - \gamma t}}}}{{\left( {{\gamma ^2} - {\Gamma ^2}} \right)}}\left( {\left( {{\gamma ^2} - {\Gamma ^2}} \right)\left( {{e^{\gamma t}} - \alpha {e^{ - \gamma t}}} \right) - \alpha \delta } \right)} \right)\notag\\& - S_2\left( \rho _B \right) \max\left\{ {{L_{11}}^2,{L_{22}}^2,{L_{33}}^2} \right\},
\end{align}
\end{widetext}
where, the eigenvalues ${\zeta _{i}}$ of the density matrix ${{\rho _{AB}}}$ are
\begin{widetext}
\begin{eqnarray}
\zeta _{1,2} &=& \frac{{{e^{ - \gamma t}}}}{{2\left( {{\gamma ^2} - {\Gamma ^2}} \right)}}\left[
\left( {{\gamma ^2} - {\Gamma ^2}} \right){e^{\gamma t}} - 2\alpha \delta \pm \sqrt {{{\left( {\left( {{\gamma ^2} - {\Gamma ^2}} \right)\left( {2\alpha {e^{ - \gamma t}} - {e^{\gamma t}}} \right) + 2\alpha \delta } \right)}^2} + 4\alpha \left( {1 - \alpha } \right){{\left( {{\gamma ^2} - {\Gamma ^2}} \right)}^2}}  \right], \nonumber \\
{\zeta _3} &=& \frac{{\left( {\gamma  + \Gamma } \right)\alpha {e^{ - \gamma t}}}}{{\left( {\gamma  - \Gamma } \right)}}\left( {{e^{ - \Gamma t}} - {e^{ - \gamma t}}} \right),
\hspace{3.3cm}
{\zeta _4} = \frac{{\left( {\gamma  - \Gamma } \right)\alpha {e^{ - \gamma t}}}}{{\left( {\gamma  + \Gamma } \right)}}\left( {{e^{\Gamma t}} - {e^{ - \gamma t}}} \right).
 \end{eqnarray}

 \end{widetext}

\section{Mixed State Elements}
\label{Mixed State Elements}
The classical correlation for the initial mixed state is given by
\begin{widetext}
\begin{equation}
{{\cal C}_2}\left( {{t}} \right) = \frac{{2{e^{ - 2\gamma t}}}}{{9{{\left( {{\gamma ^2} - {\Gamma ^2}} \right)}^2}}}\left[ \begin{array}{l}
9{\left( {{\gamma ^2} - {\Gamma ^2}} \right)^2}{e^{2\gamma t}} - {\left( {\left( {{\gamma ^2} - {\Gamma ^2}} \right)\left( {\alpha {e^{ - \gamma t}} + {e^{ - \Gamma t}}} \right) + \alpha \delta } \right)^2}\\
- {\left( {\left( {{\gamma ^2} - {\Gamma ^2}} \right)\left( {3{e^{\gamma t}} - \alpha {e^{ - \gamma t}} - {e^{ - \Gamma t}}} \right) - \alpha \delta } \right)^2}
\end{array} \right]\max \left\{ {{L_{11}}^2,{L_{33}}^2} \right\},
\end{equation}
with
\begin{eqnarray}
{L_{11}} &=& {L_{22}} = \frac{{2\alpha \gamma \Gamma Z - \alpha \left( {{\gamma ^2} + {\Gamma ^2}} \right)\sinh \left( {\Gamma t} \right) + \left( {{\gamma ^2} - {\Gamma ^2}} \right){e^{ - \Gamma t}}}}{{\sqrt {\left( {\left( {{\gamma ^2} - {\Gamma ^2}} \right)\left( {\alpha {e^{ - \gamma t}} + {e^{ - \Gamma t}}} \right) + \alpha \delta } \right)\left( {\left( {{\gamma ^2} - {\Gamma ^2}} \right)\left( {3{e^{\gamma t}} - \alpha {e^{ - \gamma t}} - {e^{ - \Gamma t}}} \right) - \alpha \delta } \right)} }}, \nonumber \\
{L_{33}} &=& \frac{{\alpha {{\left( {{\gamma ^2} - {\Gamma ^2}} \right)}^2}\left( {3{e^{\gamma t}} - \alpha {e^{ - \gamma t}} - 2{e^{ - \Gamma t}}} \right) - 2{\alpha ^2}\left( {{\gamma ^2} - {\Gamma ^2}} \right)\delta  - \left( {\left( {{\gamma ^2} - {\Gamma ^2}} \right){e^{ - \Gamma t}} + \alpha \delta } \right)}}{{\left( {\left( {{\gamma ^2} - {\Gamma ^2}} \right)\left( {\alpha {e^{ - \gamma t}} + {e^{ - \Gamma t}}} \right) + \alpha \delta } \right)\left( {\left( {{\gamma ^2} - {\Gamma ^2}} \right)\left( {3{e^{\gamma t}} - \alpha {e^{ - \gamma t}} - {e^{ - \Gamma t}}} \right) - \alpha \delta } \right)}}.
\end{eqnarray}

\end{widetext}
The quantum discord for this state can be written as
\begin{widetext}
\begin{align}
{\cal Q}\left( {{t}} \right) &= - \frac{{2{e^{ - \gamma t}}}}{{3\left( {{\gamma ^2} - {\Gamma ^2}} \right)}}\left( {\left( {{\gamma ^2} - {\Gamma ^2}} \right)\left( {\alpha {e^{ - \gamma t}} + {e^{ - \Gamma t}}} \right) + \alpha \delta } \right){\log _2}\left[ {\frac{{{e^{ - \gamma t}}}}{{3\left( {{\gamma ^2} - {\Gamma ^2}} \right)}}\left( {\left( {{\gamma ^2} - {\Gamma ^2}} \right)\left( {\alpha {e^{ - \gamma t}} + {e^{ - \Gamma t}}} \right) + \alpha \delta } \right)} \right] -\notag\\& \frac{{2{e^{ - \gamma t}}}}{{3\left( {{\gamma ^2} - {\Gamma ^2}} \right)}}\left( {\left( {{\gamma ^2} - {\Gamma ^2}} \right)\left( {3{e^{\gamma t}} - \alpha {e^{ - \gamma t}} - {e^{ - \Gamma t}}} \right) - \alpha \delta } \right){\log _2}\left[ {\frac{{2{e^{ - \gamma t}}}}{{3\left( {{\gamma ^2} - {\Gamma ^2}} \right)}}\left( {\left( {{\gamma ^2} - {\Gamma ^2}} \right)\left( {3{e^{\gamma t}} - \alpha {e^{ - \gamma t}} - {e^{ - \Gamma t}}} \right) - \alpha \delta } \right)} \right]\notag\\& + \sum\limits_{i = 1}^4 {{\zeta _i}} {\log _2}\left( {{\zeta _i}} \right)- {{\cal C}_2}\left( {{ t}} \right),
\end{align}
\end{widetext}
with the eigenvalues  ${\zeta _{i}}$ as,
\begin{widetext}
\begin{eqnarray}
{\zeta _1} &=& \frac{\alpha }{3}{e^{ - 2\gamma t}}, \hspace{0.2cm}
{\zeta _2} = \frac{{{e^{ - \gamma t}}}}{{3\left( {{\gamma ^2} - {\Gamma ^2}} \right)}}\left( {\left( {{\gamma ^2} - {\Gamma ^2}} \right)\left( {3{e^{\gamma t}} - \alpha {e^{ - \gamma t}} - 2{e^{ - \Gamma t}}} \right) - 2\alpha \delta } \right),  \nonumber \\
{\zeta _3} &=& \frac{{{e^{ - \gamma t}}}}{{3\left( {{\gamma ^2} - {\Gamma ^2}} \right)}}\left( {2\left( {{\gamma ^2} - {\Gamma ^2}} \right){e^{ - \Gamma t}} + \alpha \delta  + 2\gamma \alpha \Gamma Z - \alpha \left( {{\gamma ^2} + {\Gamma ^2}} \right)\sinh \left( {\Gamma t} \right)} \right),    \nonumber \\
{\zeta _4} &=& \frac{{\alpha {e^{ - \gamma t}}}}{{3\left( {{\gamma ^2} - {\Gamma ^2}} \right)}}\left( {\delta  - 2\gamma \Gamma Z + \left( {{\gamma ^2} + {\Gamma ^2}} \right)\sinh \left( {\Gamma t} \right)} \right). \nonumber \\
\end{eqnarray}
\end{widetext}

\section*{Acknowledgements}
This work is supported by the IET under the A F Harvey Engineering Research Prize
and funded by FCT/MEC through national funds and when applicable co-funded by FEDER-PT2020 partnership agreement under the project UID/EEA/50008/2019. One of the authors (M. I. S.)  acknowledges the CeFEMA Group for the hospitality, and providing the working conditions. H. T. thanks the support from Fundação para a Ciência e a Tecnologia (FCT-Portugal) through the Grant No. IF/00433/2015.

\end{document}